\documentclass{article}
\usepackage{spconf,amsmath,graphicx, subcaption, cite}
\usepackage{enumitem}

\title{Closing the sim-to-real gap in guided wave damage detection with
adversarial training of variational auto-encoders}
\name{Ishan D. Khurjekar, Joel B. Harley \thanks{This research is funded by the National Science Foundation under award number EECS-1839704. \newline This work has been accepted to IEEE ICASSP 2022. \newline \copyright 2022 IEEE.  Personal use of this material is permitted. Permission from IEEE must be obtained for all other uses, in any current or future media, including reprinting/republishing this material for advertising or promotional purposes, creating new collective works, for resale or redistribution to servers or lists, or reuse of any copyrighted component of this work in other works.
}}
\address{University of Florida \\
Dept. of Electrical and Computer Engineering\\
Gainesville, FL 32608}
\begin{document}
%
\maketitle
\begin{abstract}
Guided wave testing is a popular approach for monitoring the structural integrity of infrastructures. We focus on the primary task of damage detection, where signal processing techniques are commonly employed. The  detection performance is affected by a mismatch between the wave propagation model and experimental wave data. External variations, such as temperature, which are difficult to model, also affect the performance. While deep learning models can be an alternative detection method, there is often a lack of real-world training datasets. In this work, we counter this challenge by training an ensemble of variational autoencoders only on simulation data with a wave physics-guided adversarial component. We set up an experiment with non-uniform temperature variations to test the robustness of the methods. We compare our scheme with existing deep learning  detection schemes and observe superior performance on experimental data.
\end{abstract}
\begin{keywords}
Damage detection, guided waves, sim-to-real, variational auto-encoder, adversarial training
\end{keywords}
\section{Introduction}
\label{sec:intro}

Infrastructures with daily utility, such as airplanes, bridges, and buildings, need to be monitored regularly for their structural integrity. Guided wave based testing (GWT) is a popular approach for monitoring structural health as these waves can travel over long distances and are sensitive to structural defects \cite{rose2014ultrasonic}. A GWT setup consists of a spatially distributed sensor array, that can transmit and receive waves, placed on the structure to be investigated for the presence of damage. 

In this paper, we focus on the task of structural damage detection using a GWT setup. A number of target detection schemes, such as matched filtering \cite{robey1992cfar}, energy detectors \cite{urkowitz1967energy}, among others, can be applied to damage detection in ideal conditions. On the other hand, external variations, such as temperature affect wave amplitude, phase, and velocity \cite{raghavan2008effects, konstantinidis2006temperature}. This leads to a mismatch between the theoretical wave propagation model and the experimental data. For methods based on matched filtering, the model mismatch is a major challenge \cite{melvin2000space}. In addition, there is no method for perfectly removing temperature effects from data sets \cite{douglass2018dynamic}. Hence, we cannot remove temperature from the problem. 

Researchers have also proposed machine learning based approaches for monitoring structural integrity \cite{melville2018structural, khurjekar2019deep}. Yet, obtaining real-world guided wave datasets for training machine learning models is resource intensive, particularly for damage detection. While the problem can be framed as a binary classification problem,  it is difficult to obtain training data that is representative of both damage and no-damage classes.

Instead, in this paper we pose the damage detection problem as an out-of-distribution (OoD) detection problem. Generative models are commonly used for OoD detection. These include strategies based on likelihood models \cite{bishop1994novelty}, auto-regressive networks, \cite{rushe2019anomaly}, adversarial networks\cite{hou2020mahalanobis}, and variational autoencoders (VAE) \cite{sundar2020out}, among others.
A possible strategy that uses a likelihood model would be to learn a model for a related task, such as damage localization, and then threshold the likelihood value for damage detection. Yet, such an approach lacks robustness as it cannot effectively capture the input variability. Indeed, researchers have shown that generative methods are not necessarily robust as they assign spurious likelihood values to OoD inputs \cite{nalisnick2018deep}.

We propose a VAE ensemble approach  with two salient features to enhance applicability to realistic guided wave damage detection:
\begin{itemize}[itemsep=1mm, parsep=0pt]
\item We train the VAE ensemble on simulation data alone, eliminating the need to set up resource intensive experiments for training data generation. 
\item We simulate wave physics-based adversarial perturbations in the training data to enable robustness to input variability (temperature variations).\end{itemize} 
We choose the VAE objective value (lower bound on the data likelihood) \cite{kingma2013auto} as the damage detection statistic since large values indicate that we are in-distribution and therefore match the simulation data. We compare the performance of our approach with other deep generative approaches on experimental data with non-uniform temperature variations. Our approach achieves superior detection performance and well-separated statistic values signifying superior robustness 

\section{Sim-to-real damage detection framework}
We propose a VAE framework for detecting the presence of damage in a structure using guided waves with two salient features: training on simulation data alone with simulated adversarial perturbations for  robustness to temperature variations. The problem setup and the framework is explained in the following subsections.


\begin{figure}[!t]
    \centering
    \includegraphics[width = 3.2in]{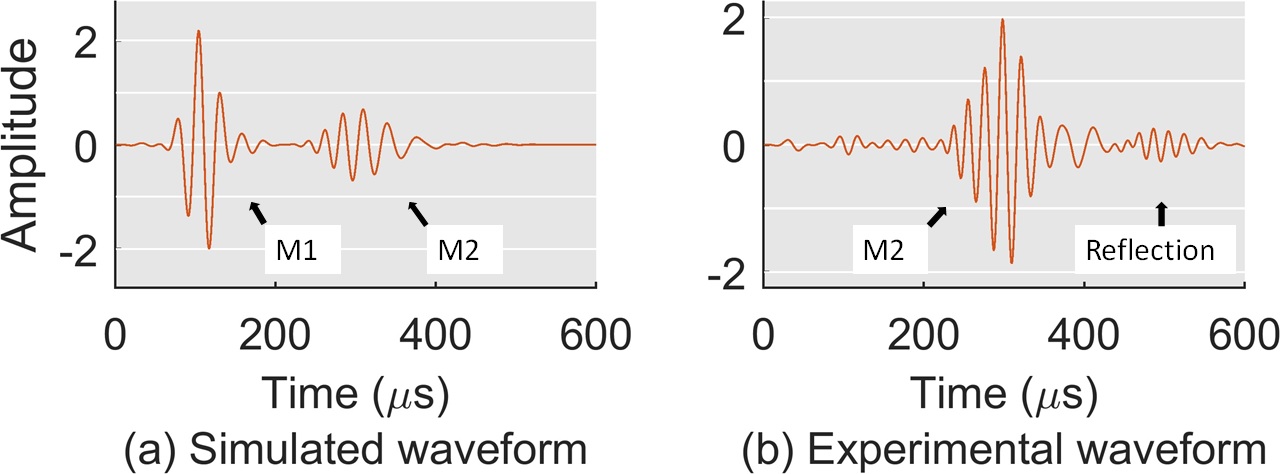}
    \caption{Simulation and experimental waveform comparison. Both the waves travel a distance of 0.5681m}
    \label{fig:simexpt_comp}
    \vspace{-3mm}
\end{figure}
\subsection{Guided wave setup}
In this paper, we simulate guided wave structural health monitoring data. We assume the structure to be a square plate that is investigated for the presence of damage. A sensor array placed on the structure transmits and receives signals (wideband signals with $Q$ frequencies and $M$ sensor pair measurements). A Lamb wave model is used to describe the wave propagation \cite{lamb1917waves}. The general Lamb wave model is given as,
\begin{equation}\label{lambwave} 
    x(\omega,r) = \sum_{n}\sqrt \frac{1}{\kappa_{n}(\omega)r}s(\omega) e^{-j\kappa_{n}(\omega)r},
\end{equation}
where the guided wave signal, $x(\omega, r)$, for frequency $\omega$ and at distance $r$ from source is modeled as a superposition of wave modes. $s(\omega)$ is the transmitted signal and $\kappa_{n}(\omega)$ is the frequency and mode dependent wavenumber. 

We assume the received signal travels two paths. The first path is directly from the transmitter to the receiver (baseline signal: $x_{b}$). The second path is from the transmitter to the damage and then to the receiver (damage signal: $x_{s}$). This is mathematically expressed as 
\begin{equation}
\label{wave_damage}
    x(\omega, r) = x_{b}(\omega, r) + \alpha x_{s}(\omega, r),
\end{equation}
where $\alpha$ is the reflection coefficient. As we standardize data, the choice of $\alpha$ does not affect the results. 

Fig.~\ref{fig:simexpt_comp}(a-b) shows the simulated and experimental guided wave signals respectively. Note that the wave mode-2 (referred to as A0 mode \cite{lamb1917waves}) coincides in the simulated and experimental signal but mode-1 (referred to as S0 mode \cite{lamb1917waves}) is weak in the experimental signal. 

Before processing the data, we apply baseline subtraction. That is, the baseline signal ($x_{b}$) is subtracted in order to isolate the damage signal ($x_{s}$). Ideally, only the damage signal and noise should remain after baseline subtraction. Yet, baseline subtraction is not perfect in presence of distorting effects of temperature variations as we describe in Section~\ref{baseline_sub}. While methods exist to reduce the effects of temperature \cite{harley2012scale}, no method can perfectly remove them.

\subsection{Variational autoencoder: VAE}
\begin{figure}[!t]
    \centering
    \includegraphics[width = 2.8in]{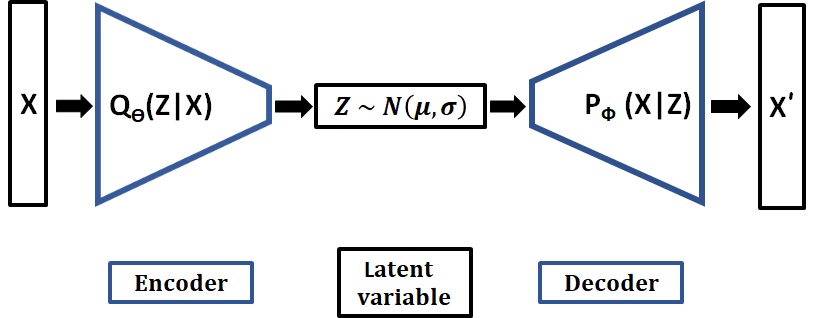}
    \caption{VAE architecture}
    \vspace{-3mm}
    \label{fig:vae_arch}
\end{figure}
We pose the damage detection problem as an OoD detection problem for which we use a VAE. We train the VAE on the baseline subtracted signal with the assumption of damage as in (\ref{wave_damage}). Hence signals without damage component are out-of-distribution. VAE consists of an encoder and a decoder as shown in Fig.~\ref{fig:vae_arch}. It reconstructs the input data using a probabilistic latent variable model. Specifically, the lower bound on data likelihood (ELBO: evidence lower bound) \cite{kingma2013auto} is maximized. The ELBO is written as,
\begin{equation}
\label{elbo}
    \log p(x) \geq E_{z\sim Q_{\theta}}[\log P_{\phi}(x|z)] - \mathcal{D}[Q_{\theta}(z|x) || P_{\phi}(z)],
\end{equation}
where $Q_{\theta}$ and $P_{\phi}$ are the encoder and decoder networks, respectively.
The first term in (\ref{elbo}) is the cross entropy and the second term is the Kullback-Liebler divergence. The latent random variable $z$ is assumed to follow a Normal distribution.

\subsection{Closing the sim-to-real gap}
Researchers have built robust deep learning models by training on data with adversarial perturbations \cite{wang2018unsupervised, gokhale2020attribute}. Therefore, we simulate adversarial perturbations in the training data based on empirical modeling of temperature effect on wave propagation. These perturbations are simulated by multiplying the wavenumber in (\ref{lambwave}) by a random factor defined as, 
\begin{equation}
\label{wavenumber_distort}
    \kappa_{n}'(\omega) = \gamma \kappa_n(\omega),
\end{equation}
where $\gamma$ is the random multiplicative factor sampled uniformly from the interval $[1-\delta, 1 + \delta]$. We choose $\delta = 0.02$ to match the range of wavenumber variations ($\pm 2 \%$) in the experimental data as caused by temperature. 
Further, we train an ensemble ($n = 10$) of VAE's each with a different weight initialization, as ensembling  has also been shown to increase robustness on acoustic tasks \cite{noman2019short}.
\begin{figure*}[!h]
\centering
    \includegraphics[width = 5.75in]{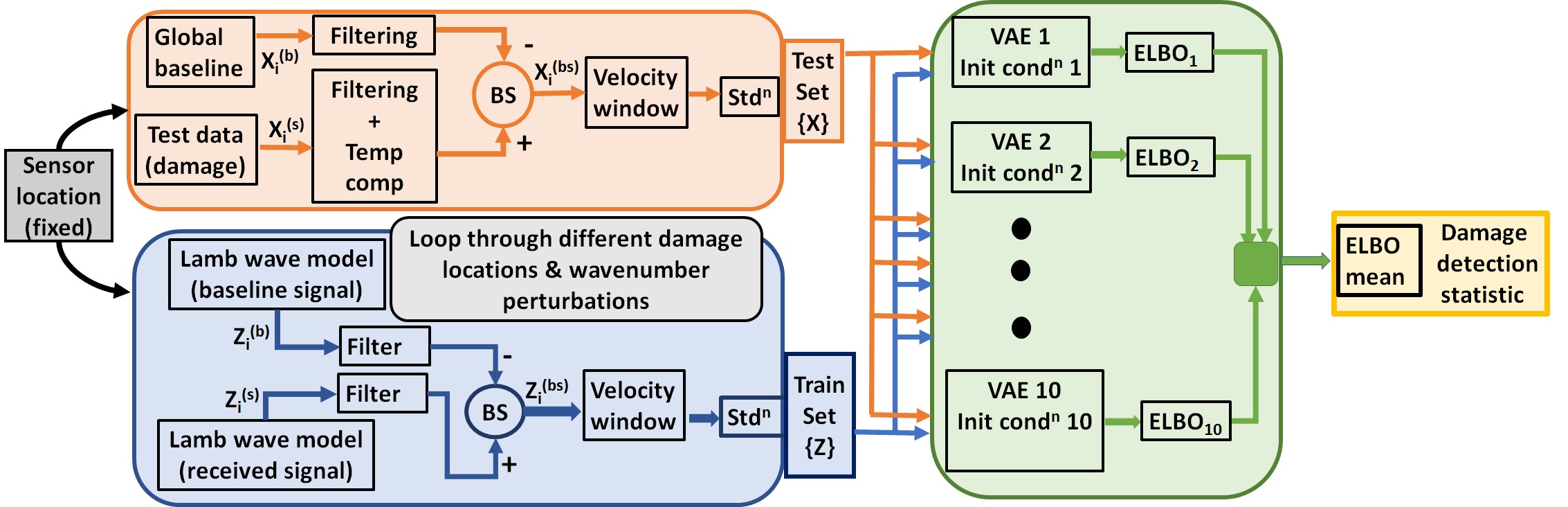}
    \caption{Framework validation setup}
    \label{fig:damagedetection_framework}
     \vspace{-3mm}
\end{figure*}

\subsection{Damage detection}
We define the damage detection statistic ($\tau$) as,
\begin{equation}\label{bpd}
    \tau(x) = \frac{1}{Q \times M}\frac{1}{n}\sum_{i=1}^{n}{ELBO_{i}(x)},
\end{equation}
where $\tau(x)$ is the mean of ELBO estimates ($\textrm{ELBO}_{i}$'s) from the VAE ensemble normalized for $Q$~frequencies and $M$~sensor pair measurements. Intuitively, we should have high statistic value when the test sample is ``in-distribution'' (damage) and conversely low statistic value when test sample is ``out-of-distribution'' (no damage).

\section{Framework validation}
The complete damage detection framework is illustrated in Fig.~\ref{fig:damagedetection_framework}. The individual framework components are explained in the subsequent subsections.
\label{sec:motivation}

\subsection{Implementation}
\label{sim_setup}

For training, we simulate $Q = 1000$ frequencies and $M = 240$ sensor pair measurements (16 element sensor array) to match the experimental setup, described in Section~\ref{expt_setup}.
We refer to one spatio-temporal observation matrix of dimensions $Q \times M$ as one sample. We simulate $t = 5000$ samples with 4000 samples used as training set and the rest 1000 samples used as the validation set. We convert samples to time-domain and standardize them before inputting to VAE. 

The wave signal is pulse compressed to remove unwanted dependence on the transmitted signal phase. We remove the initial 40 $\mu$s of signal to remove electromagnetic interference. We pass the signal through a low pass Gaussian filter (center frequency $f_c = 37.5$~kHz and bandwidth $B = 30$~kHz) as the effects of damage are observed at these lower frequencies. We apply an exponentially tapering velocity window ($v_{win} = 1500$~m/s) to remove the unwanted reflections from the plate boundaries. 

The VAE architecture is detailed in Table~\ref{vae_init}. We use 1D convolutional layers in the encoder and transposed 1D convolutional layers (also called deconvolution) in the decoder. We apply batch normalization after every layer for faster training. We apply dropout regularization for all dense layers with dropout probability = 0.1. We use the reparameterization trick for latent space sampling \cite{kingma2013auto}. We train each VAE in the ensemble with simulated data alone for 15 epochs with a batch size of 16. All the above mentioned parameter values are chosen to maximize detection performance.
\begin{table}[!h]
\caption{VAE architecture}
\begin{tabular}{ p{1.6cm} p{4.6cm} p{1.25cm} }
 \hline
 Layer  &Layer description &Activation\\
 \hline
 Conv1D &   Filters = 12 ; kernel size = 3 ; stride length = 2 & ReLU\\
  \hline
 Conv1D & Filters = 24; kernel size = 3 ; stride length = 2 & ReLU\\
 \hline
  Dense & Fully connected ; nodes = 1200 & Sigmoid\\
  \hline
  Dense $\times 2$ & latent\_dim = 2& -\\
  \hline
  Dense&   Fully connected ; nodes = 1200  & Sigmoid\\
  \hline
 Dense&   Fully connected ; nodes = Q$\times$M  & Sigmoid\\
  \hline
 Conv1D Tanspose &  Filters = 24; kernel size = 3 ; stride length = 2 & ReLU\\
 \hline
 Conv1D Tanspose & Filters = 12; kernel size = 3 ; stride length = 2 & ReLU\\
 \hline 
\end{tabular}
\label{vae_init}
 \vspace{-3mm}
\end{table}

\subsection{Baseline subtraction}
\label{baseline_sub}
The baseline signal has to be subtracted from the received signal to isolate the damage signal while compensating for the effect of temperature. Popular strategies include choosing from a baseline bank \cite{lu2005methodology} and / or stretching signals using scale transform \cite{harley2012scale}. We create a calibration signal bank by choosing one signal each from the damaged and undamaged experimental signal set randomly. The calibration bank is considered as an extension of the validation set. For a particular test signal, we choose the calibration signal that minimizes residual energy. We stretch the test signal using the scale transform \cite{harley2012scale} (removing some of the effects caused by temperature) to match the chosen calibration signal and then subtract a globally chosen baseline signal.

\subsection{Experimental setup}
\label{expt_setup}

\begin{figure}[!t]
    \centering
     \includegraphics[ width = 2.8in]{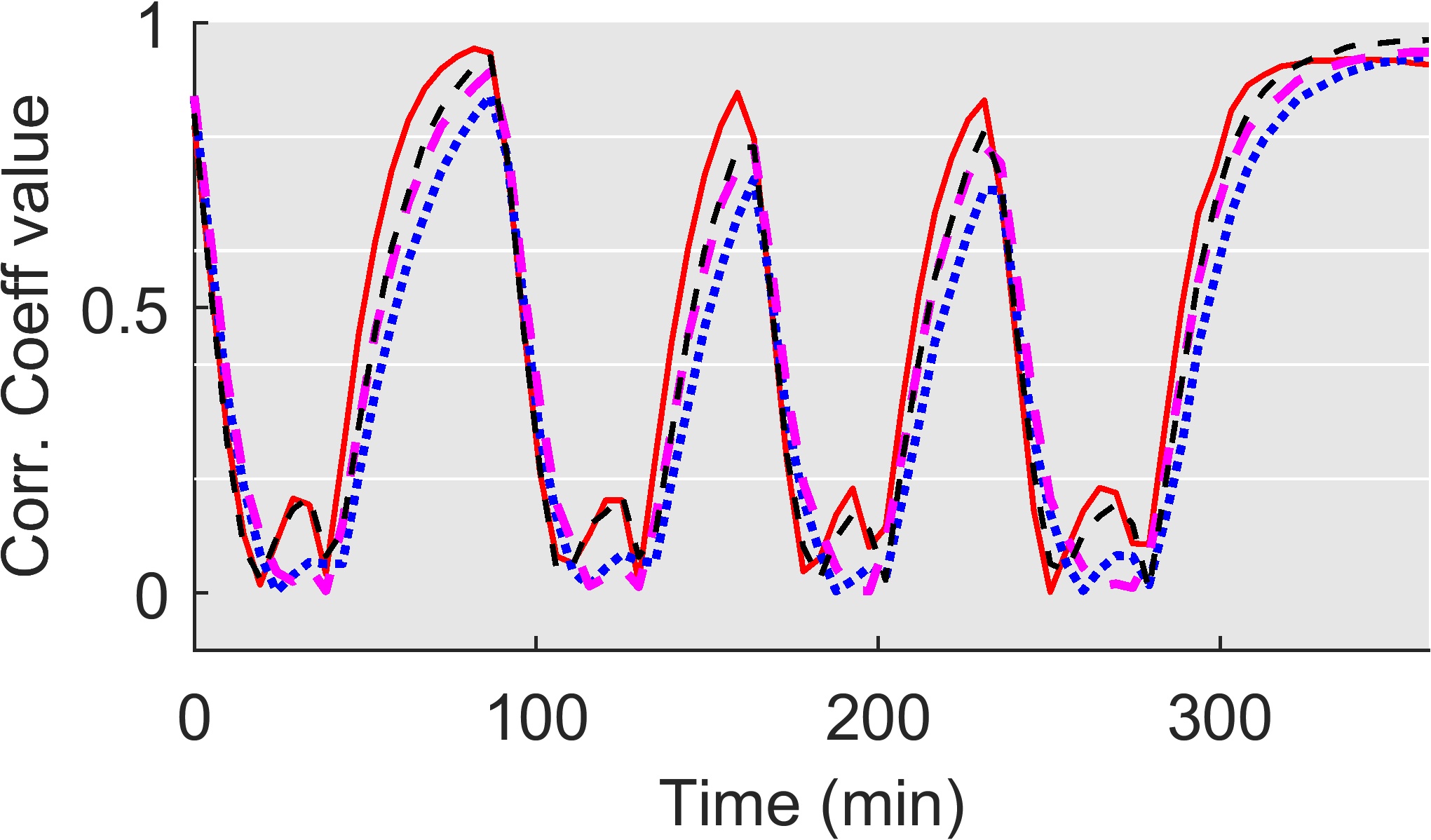}
    \caption{Correlation between first experimental measurement and successive measurements for 4 sensor pairs}
    \label{fig:expt_var}
    \vspace{-2mm}
\end{figure}

We mount 16 sensors at random locations on an aluminum plate of size $1.22$ m $\times$ $1.22$ m. We set up the data acquisition system to have a sampling rate of 1 MHz. We transmit a 0.1 ms long chirp signal with a frequency sweep of 50 kHz to 500 kHz.  We record 4 ms long measurements on the receiving sensors. Spatio-temporal variations are introduced over the plate using a heating fan placed in a corner. 
The temperature is varied periodically from approximately 24$^{\circ}$C to 39$^{\circ}$C across the plate. We collect 76 measurements over $\approx{6}$ hours. We physically simulate damage by placing a mass at $(0.53,0.60)$ m from the $37^{th}$ measurement onward. Fig.~\ref{fig:expt_var} shows the correlation values between first and successive measurements. This illustrates the effect of temperature variations on wave propagation.

\section{Results}
\label{sec:typestyle}
We compare our VAE based scheme with a likelihood model-based detection scheme, utilizing a feedforward neural network. We train a network with guided wave data generated using (\ref{lambwave}) to identify corresponding damage locations. This network uses a Gaussian likelihood as the objective function, which is used as a damage detection metric. 

The detection threshold(s), ($\tau_{0}$), for the method(s) are computed individually as follows: The detection statistic values are calculated for the same randomly chosen calibration signals used for baseline subtraction. The mid-point of these values is chosen as the detection threshold. This is done to maximize separation between damage and no damage case as well as for fair comparison of all methods. We define the probability of damage detection ($p_{d}$) from the detector as
\begin{equation}
    p_{d} = p(~\tau\geq \tau_{0}~|~H_{A}),
\end{equation} and the probability of false alarm ($p_{fa}$)  as
\begin{equation}
    p_{fa} = p(~\tau\geq \tau_{0}~|~H_{0}),
\end{equation}
where $\tau$, $\tau_{0}$, $H_{0}$, $H_{A}$ represent the detection statistic, detection threshold, null hypothesis (no damage), and alternative hypothesis (damage) respectively.

\begin{figure}[!t]
    \centering
    \includegraphics[width = 3.0in]{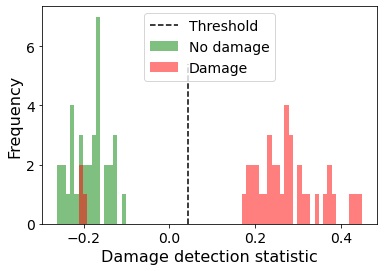}
    \caption{VAE ensemble damage detection histogram}
    \label{fig:elbo_hist}
    \vspace{-3mm}
\end{figure} 

\begin{table}[!t]
\caption{Damage detection performance}
 \centering
\begin{tabular}{ p{4.5cm} p{0.9cm} p{0.9cm} }
 \hline

 Method  &$p_{d}$ &$p_{fa}$\\
 \hline
VAE-adv training ($*$)&   0.923 & 0.000\\
  \hline
 VAE-ideal training &   0.461 & 0.000\\
 \hline
  Likelihood model-adv training &   0.923 & 0.600\\
  \hline
 Likelihood model-ideal training &   0.487 & 0.712\\
  \hline
\end{tabular}
\label{perf_comp}
 \vspace{-2mm}
\end{table}

Table~\ref{perf_comp} shows the performance comparison of the VAE based approach and the likelihood model-based approach. We also compare the performance of models trained on ideal data and on data with adversarial perturbations (denoted as -ideal training and -adv training respectively). We first observe that models trained on data with adversarial perturbations have superior performance compared to those trained on ideal data. This underscores the importance of adversarial training for enhancing robustness. 

Next, we compare the performance of our approach and the likelihood model-based detection scheme. The probability of detection ($p_{d}$) is equal for both but our VAE scheme has a much better false alarm rate ($p_{fa}$: 0.00 compared to 0.60). This is in line with the observation that likelihood models assign spurious likelihood values to OoD inputs. Fig~\ref{fig:elbo_hist} shows the histogram of the damage detection statistic for our scheme (VAE trained on adversarial data). The detection statistic is well-separated, signifying robustness.

\section{Conclusions}
Here, we pose the problem of guided wave-based damage detection problem as an OoD detection problem. We propose a VAE ensemble network based approach for this task which is trained on simulation data alone. Ensemble approach together with adversarial training enables robustness. To validate the framework, we set up an experiment to collect guided wave data in presence of non-uniform temperature variations. Results illustrate that the detection performance of the proposed framework is robust to temperature variations and superior to other deep generative methods.

\label{sec:majhead}



\bibliographystyle{IEEEbib}
\bibliography{refs}

\end{document}